\documentclass[conference]{IEEEtran}
\IEEEoverridecommandlockouts
\usepackage{cite}
\usepackage{amsmath,amssymb,amsfonts}
\usepackage{graphicx}
\usepackage{textcomp}
\usepackage{xcolor}
\usepackage{amsbsy,marvosym,caption,threeparttable,subfigure,float,lastpage,lscape,listings}
\usepackage{eurosym,mathrsfs,fancyhdr,CJK,multicol,indentfirst,color,bm,upgreek,warpcol}
\usepackage{titlesec}
\usepackage{epstopdf}
\usepackage{algpseudocode}
\usepackage{graphics}
\usepackage{epsfig}
\usepackage{tabularx}
\usepackage[a4paper, total={184mm,239mm}]{geometry}
\allowdisplaybreaks[3] 				
\usepackage{bbm} 					
\usepackage{multirow} 				
\usepackage{booktabs} 				

\usepackage{enumitem}				
\usepackage{amsthm} 				
\newtheorem{owndefinition}{Definition}		
\newtheorem{ownproperty}{Property} 			
\newtheorem{ownexample}{Example} 			
\usepackage[linesnumbered,ruled,vlined]{algorithm2e}
\usepackage[numbers, sort & compress]{natbib}
\def\BibTeX{{\rm B\kern-.05em{\sc i\kern-.025em b}\kern-.08em
    T\kern-.1667em\lower.7ex\hbox{E}\kern-.125emX}}
\begin{document}

\title{A Semi-Tensor Product based Circuit Simulation for SAT-sweeping}

\author{
    \IEEEauthorblockN{Hongyang Pan\textsuperscript{1}, Ruibing Zhang\textsuperscript{2}, Yinshui Xia\textsuperscript{2}, Lunyao Wang\textsuperscript{2}, Fan Yang\textsuperscript{1} and Xuan Zeng\textsuperscript{1}, Zhufei Chu\textsuperscript{2}}
    \IEEEauthorblockA{
        \textit{\textsuperscript{1} Fudan University, \textsuperscript{2} Ningbo University} \\
        \{yangfan, zengxuan\}@fudan.edu.cn, chuzhufei@nbu.edu.cn \\
    }
    \vspace{-20pt}
}
\maketitle
\vspace{-10em}

\begin{abstract}
In recent years, circuit simulators and Boolean satisfiability (SAT) solvers have been tightly integrated to provide efficient logic synthesis and verification.
Circuit simulation can generate highly expressive simulation patterns that can either enumerate or filter out most candidates for synthesis.
Subsequently, SAT solvers are employed to check those that remain, thereby making the logic synthesis process more efficient.
This paper introduces a novel circuit simulator of $k$-input lookup table ($k$-LUT) networks, based on semi-tensor product (STP).
STP-based simulators use computation of logic matrices, the primitives of logic networks, as opposed to relying on bitwise logic operations for simulation of $k$-LUT networks.
Experimental results show that our STP-based simulator reduces the runtime by an average of 7.2$\times$.
Furthermore, we integrate this proposed simulator into a SAT-sweeping engine known as SAT sweeper.
Through a combination of structural hashing, simulation, and SAT queries, SAT sweeper simplifies logic networks by systematically merging graph vertices from input to output.
To enhance the efficiency, we used STP-based exhaustive simulation, which significantly reduces the number of false equivalence class candidates, thereby improving the computational efficiency by reducing the number of SAT calls required.
When compared to the SOTA SAT sweeper, our method demonstrates an average 35\% runtime reduction.
\end{abstract}


\vspace{-2mm}

\setlength{\arraycolsep}{1.5pt}
\renewcommand\arraystretch{0.5}
\section{Introduction}
\label{sec1}
It is common for logic synthesis and verification applications to integrate the \emph{Boolean satisfiability} (SAT) solver with a random or guided simulator.
Currently, logic optimization methods often encode problems as SAT problems to enhance optimization quality through the use of efficient SAT solvers, where simulators can assist it in reducing candidates in order to avoid time-consuming satisfiable runs. 
Additionally, the SAT solver can find a \emph{counter-example} (CE), which the simulator can use to falsify other properties to save future SAT calls~\cite{intro1}.
There are many applications that take advantage of this integration, such as SAT-sweeping.
In SAT-sweeping, the simulator can pregenerate a set of simulation patterns for a given Boolean network, most non-equivalence can be effectively eliminated by simply comparing simulation signatures.
After each satisfiable SAT call, it also simulates each CE immediately in order to disprove as many properties as possible~\cite{intro3,intro4,intro2}.

The efficiency of simulation, encompassing both initial simulation and CE simulation, plays an important role in SAT-sweeping, with the principal aim of minimizing the number of expensive NP-hard SAT solver calls.
Certain algorithms incorporate partial simulation as a strategy to expedite simulation speed. 
However, in scenarios where the pattern set is insufficiently expressive, the equivalence classes tend to enlarge, and the refinement via CE may become computationally more expensive.
In practice, the equivalence class usually contains a few percent of the total gates in a valid merge.
Ideally, we only need to simulate these gates, which must be simulated, with exhaustive patterns for better runtime.

In this paper, we propose a \emph{$k$-input lookup table} ($k$-LUT) circuit simulator based on \emph{semi-tensor product} (STP).
Modern circuit simulators support bitwise logic operations, utilizing fast bit-parallel simulation to enhance efficiency. 
Nonetheless, simulating $k$-LUT networks poses a unique challenge, as it does not readily exploit these bit-parallel capabilities. 
$k$-LUT simulation must obtain information in turn regarding each simulation pattern by traversing all nodes in a topological order before computing the output values of each node.
The STP method works by matrices, utilizing logic matrices for the definition of Boolean variables to prove the basic logic properties~\cite{intro6}. 
As primitives in the logic network, logic matrices, which convert logical reasoning into mathematical computations and preserve topological information between circuits, are used to simulate the $k$-LUT network.
Motivated by this reasoning, we integrate our STP-based simulator into the SAT-sweeping engine, where efforts are made in order to run as many simulations as possible within the given runtime budget, reducing equivalence classes to avoid unnecessary SAT solver calls.

The main contributions are: (1) a more efficient simulator: compared with other bitwise logic operation based simulation implementations, the STP-based simulator does not require specific logic operators, and the output values of any node can be computed by one matrix pass; 
(2) the integration of STP-based simulator into SAT sweeper \emph{\&fraig}: we first use SAT-guidance initial simulations to refine equivalence classes while adding the constant node substitution, which can significantly reduce the number of false candidates for merging~\cite{intro7}.
Then, map the nodes of non-equivalence classes to $k$-LUTs and simulate the nodes of equivalence classes with exhaustive simulation pattern to further refine candidate equivalence classes.
In addition, for the candidate gates to be validated in each valid merge, we consider its \emph{transitive fanin} (TFI) to explore the maximum QoR.
Experimental results show that the simulator can reduce the runtime by 7.18$\times$ on average, and the SAT sweeper can reduce the runtime by 35\% on average.

\vspace{-2mm}
\section{Preliminaries}
\label{sec2}
\subsection{Circuit simulation}
\label{sec21}
Circuit simulation involves visiting nodes in topological order and computing output values using their input values.
The \emph{simulation pattern} is a collection of Boolean values assigned to each \emph{primary input} (PI) of a network.
Practically, multiple simulation patterns can be consolidated by encoding sequences of Boolean values as machine words, rather than as individual bit. 
Modern arithmetic logic units are enabling the computation of 32 or 64 patterns for a node within a single CPU instruction.
The \emph{simulation signature} of a node is an ordered set of values produced at the node under each simulation pattern. 
If a set of simulation patterns covers all possible combinations of value assignment (a requirement necessitating $2^k$ patterns for $k$ PIs), this set is \emph{exhaustive} and the simulation signatures are also known as \emph{truth tables} (TTs)~\cite{pri1}.

Simulation can be executed either globally across the entire network or locally in a small \emph{window} that called \emph{partial simulation}. 
$2^{16}$ patterns are already impractical to handle, however, networks typically feature a greater number of primary inputs, often exceeding 16. 
To use the exhaustive set of patterns, simulation must be restricted to a window encompassing fewer than 16 (typically within the range of 8 to 10) leaf nodes.

\subsection{Semi-Tensor Product of Matrices and Its Logical Reasoning}
\label{sec22}
This subsection gives a brief review of the STP computation of matrices.
We refer the reader to~\cite{pri2,pri21} for more details.
The real matrices with $m \times n$ dimensions are represented by $M^{m \times n}$.
Consider two matrices $X \in M^{m \times n}$ and $Y \in M^{p \times q}$, the STP can produce matrices in any dimension.

\begin{owndefinition}
\label{def1}
Let $X \in M^{m \times n}$ and $Y \in M^{p \times q}$, the STP of $X$ and $Y$, denoted by $X \ltimes Y$, is defined as
\begin{center}
$ X \ltimes Y = (X \otimes I_{t/n}) \cdot (Y \otimes I_{t/p})$,
\end{center}
where $\cdot$ represents the common matrix product, $I_n$ represents the identity matrix with dimension $n$, $t$ is the least common multiple of $n$ and $p$, and $\otimes$ is Kronecker product of two arbitrary dimensional matrices~\cite{pri3}.
\end{owndefinition}

\begin{ownproperty}
\label{pro1}
The STP of matrices supports matrix swapping.
Let $A$ be a matrix with any dimensions, if $Z_r \in M^{1 \times t}$ is a row vector, then $A \ltimes Z_r = Z_r \ltimes (I_t \otimes A)$.
In contrast, if $Z_c \in M^{t \times 1}$ is a column vector, then $Z_c \ltimes A = (I_t \otimes A) \ltimes Z_c$.
\end{ownproperty}

The matrix form of logic formulas can be used to describe logic representations in general.
We refer to the matrix product as the STP in this paper and omit the symbol ``$\ltimes$'' hereinafter.
First, we denote the set of Boolean variables $\mathbb{B}$.
\begin{equation}
\label{eq1}
\mathbb{B} : \left\{True = \begin{bmatrix} 1 \\ 0 \end{bmatrix} , False =\begin{bmatrix} 0 \\ 1 \end{bmatrix}. \right\}
\end{equation}

\begin{owndefinition}
\label{def2}
A $M^{2 \times 2^n}$ matrix is called a logic matrix if all its columns are elements in $\mathbb{B}$, where logic matrix $M_{\sigma}$ in which columns are consistent with the TT (it is read from right to left) of a logic operation $\sigma$ is called the structural matrix.
\end{owndefinition}

\begin{ownproperty}
\label{pro2}
$a, b \in \mathbb{B}$ and $\sigma$ is an any Boolean operator.
The structural matrix of unary operator ``\emph{not}'' ($\lnot$) is $M_{\lnot} = \begin{bmatrix} 0 & 1 \\ 1 & 0 \end{bmatrix}$.
The inversion of variable $a$ can be converted to matrices multiplication as $\bar{a} = M_\lnot a$.
Similarly, for binary operators, the logic representation can be converted as $a\ \sigma\ b = M_{\sigma} ab$.
\end{ownproperty}

Therefore, any Boolean function can be converted into its \emph{STP form} by structural matrices, and logic identities can be easily proved using structure matrices of Boolean operators and STP properties.

\begin{ownexample}
Prove the logic identity $a \rightarrow b = \bar{a} \lor b$ using the STP based computation.
\begin{proof}
According to the Property~\ref{pro2}, the STP form of the left hand side is $M_{\rightarrow} ab$, while the right hand side is $M_{\lor} (M_\lnot a)b$.
\begin{center}
$M_{\lor} M_\lnot = \begin{bmatrix} 1&1&1&0 \\ 0&0&0&1 \end{bmatrix} \begin{bmatrix} 0&1 \\ 1&0 \end{bmatrix} = \begin{bmatrix} 1&0&1&1 \\ 0&1&0&0 \end{bmatrix} = M_{\rightarrow}$.
\end{center}
Hence, the identity holds.
\end{proof}
\end{ownexample}

A key aspect of Boolean function manipulation is the \emph{canonical form} (CF), since these functions can be functionally equivalently represented in several different logic realizations. A canonical form is also available for STP.

\begin{ownproperty}
\label{pro3}
Any logic expression $\Phi(x_1, \ldots, x_n)$ with Boolean variables $x_1, \ldots, x_n \in \mathbb{B} $ can be computed into a canonical form $M_{\Phi}$ as
\begin{center}
$ \Phi(x_1, \ldots, x_n) = M_{\Phi} x_1 \ldots x_n$,
\end{center}
where $M_{\Phi} \in  M^{2 \times 2^n}$.
\end{ownproperty}

We use an example to explain the STP computation process.

\begin{ownexample}
\label{exm2}
There are three persons $a$, $b$, and $c$. They are either honest or liar, suppose a liar always said a lie and the honest man always told the truth.
Person $a$ said that person $b$ is a liar, person $b$ said person $c$ is a liar, and person $c$ said that both $a$ and $b$ are liars. Who is/are the liar(s)?

First, we define logic variable $a$ to indicate person $a$ is honest. 
Thus $\bar a$ means $a$ is a liar. 
The definitions also work for Boolean variables $b$ and $c$.
The statements result in the logic expression
\begin{equation}
\label{liar}
\Phi(a, b, c) = (a \leftrightarrow \bar b) \land (b \leftrightarrow \bar c) \land (c \leftrightarrow \bar a \land \bar b).
\end{equation}
The STP form of~\eqref{liar} is
\begin{center}
  $\Phi = M_{\land}^{2}(M_{\leftrightarrow} a M_\lnot b)(M_{\leftrightarrow} b M_\lnot c)(M_{\leftrightarrow} c M_{\land} M_\lnot a M_\lnot b)$.
\end{center}
Then, converting the STP form of logic expression into the canonical form $M_{\Phi}$ as
\begin{center}
    $\Phi(a, b, c)  = M_{\Phi}abc  = \begin{bmatrix} 0&0&0&0&0&1&0&0 \\ 1&1&1&1&1&0&1&1  \end{bmatrix} abc.$
\end{center}

If we assume the simulation pattern is 010, that is, $b$ is honest, $a$ and $c$ are liars,
\begin{center}
$a =\begin{bmatrix} 0 \\ 1  \end{bmatrix} , b =\begin{bmatrix} 1 \\ 0  \end{bmatrix} , c =\begin{bmatrix} 0 \\ 1  \end{bmatrix}.$
\end{center}
The STP form $\Phi(a, b, c)$ can be computed, i.e., simulated as 
\begin{align}
  \Phi(a, b, c) & = \begin{bmatrix} 0&0&0&0&0&1&0&0 \\ 1&1&1&1&1&0&1&1  \end{bmatrix} \begin{bmatrix} 0 \\ 1  \end{bmatrix}\begin{bmatrix} 1 \\ 0  \end{bmatrix}\begin{bmatrix} 0 \\ 1  \end{bmatrix} \notag  \\
                & = \begin{bmatrix} 0&1&0&0 \\ 1&0&1&1  \end{bmatrix} \begin{bmatrix} 1 \\ 0  \end{bmatrix}\begin{bmatrix} 0 \\ 1  \end{bmatrix} \notag \\
                & = \begin{bmatrix} 0&1 \\ 1&0  \end{bmatrix} \begin{bmatrix} 0 \\ 1  \end{bmatrix} = \begin{bmatrix} 1 \\ 0  \end{bmatrix}.  \notag
\end{align}
\end{ownexample}

\subsection{SAT-sweeping}
\label{sec23}
SAT-sweeping is used to detect, prove, and merge (or collect) functionally equivalent nodes (up to complementation). 
In SAT-sweeping, two nodes are checked if they can be merged using SAT~\cite{pri4,sweep2}. 
SAT solvers provide a CE in the event the nodes cannot be merged, which is an input assignment that allows the two gates to be simulated to have different values.
A traditional implementation of SAT-sweeping would test all possible pairs of nodes.
In order to alleviate this problem, simulation is extensively used in SAT-sweeping in order to reduce the number of calls to the SAT solver.
Using initial random simulations, nodes can be grouped into equivalence classes, that is, classes of nodes that always simulate to the same value.
In this case, only calls to SAT are required to prove, or disprove, equivalencies between gates belonging to the same class.
As a result, the number of SAT queries has already been drastically reduced~\cite{sweep1}. 

\vspace{-2mm}
\section{STP-based circuit simulator}
\label{sec3}
In this section, we propose a STP-based simulation of $k$-LUT networks.
In modern simulator, the key aspect of simulating $k$-LUTs is applying bitwise logic operations efficiently. 
A $k$-LUT takes $k$-input bits, and the simulator applies bitwise logic operations to its input signals, simulates the behavior by a predefined TT of LUTs.
However, these bitwise operations (AND, OR, XOR, and NOT) can not provide efficient support for $k$-LUT networks, making simulation slower.
When it comes to STP, any Boolean function can be easily converted into its $k$-LUT network (bitwise operation is 2-LUT) represented by matrices, and the simulation is actually matrix multiplication.

\subsection{Circuit Simulation Algorithm}
\label{sec31}
The STP-based simulator can simulate all nodes or some specified nodes.
The former visits all nodes in topological order, while the latter use the cut algorithm in Section~\ref{sec33} to map these nodes which do not simulated into $k$-LUT. 

The algorithm is shown in Algorithm~\ref{algo1}.
The input of the algorithm is a $k$-LUT network \emph{K}, a simulation pattern set \emph{P} and the simulation mode \emph{m}.
The simulation mode contains all node simulation ($a$) or specified node simulation ($s$).
The output is the obtained simulation signature \emph{S} according to the choice of \emph{m}.
If the chosen mode is $a$, the simulator will visit all nodes in the network in a topological order and use their input value to compute the output value by matrices multiplication(line 2).
Otherwise, when we only need to get the simulation signature of specified nodes, we perform the simulation by following the steps below.
First, we compute the size \emph{limit} of a cut based on the number of simulation patterns.
Because it also takes time to compute the TT of cuts, this ensures that the STP method is more efficient than direct simulation(line 4).
Second, we take nodes in $s$ and \emph{limit} as the boundary to cut \emph{K}, so that each cut is a tree structure with leaf node no larger than \emph{limit}, and the root node of each cut is stored in \emph{root} set(line 5).
Then, we use the STP-based matrices multiplication to compute the TT of all cuts in the \emph{root} set(line 6).
Finally, each cut node in the root collection is accessed in topological order and its output is computed based on the input value, as shown in Example~\ref{exm2} (line 7).
The algorithm returns the desired simulation signatures.

\begin{algorithm}[t]\footnotesize
  \SetAlgoLined
  \LinesNumbered 
  \caption{STP-based circuit simulation}
  \label{algo1}
  \KwIn{logic network \emph{K}, simulation pattern \emph{P}, \emph{m}(all nodes \emph{a} or specified nodes \emph{s})}
  \KwOut{simulation signature(\emph{S})}
  \eIf{m == $a$}{
    $S \leftarrow$ sim\_all\_nodes($K,P$)\;
    \textbf{return} $S_a$\;
  }{
    $n \leftarrow P$.size(), \emph{limit}=\emph{log(n)} \; 
    $root \leftarrow$ circuit\_cut($K,limit,s$)\;
    STP\_matrices\_multiplication($K,root$)\;
    $S \leftarrow$sim\_nodes($K,P$)\;
    \textbf{return} $S_s$\;
  }
\end{algorithm}

\subsection{Cut Algorithm}
\label{sec33}
The conventional method of circuit simulation involves a traversal of all nodes in the circuit, following a topological order, and computing their output values based on input values. 
However, this method becomes redundant when our objective is to obtain the simulation signature for specific nodes, rather than all nodes.
To address this issue, we introduce a novel cut algorithm designed to minimize the number of intermediate nodes that require simulation. 
This optimization reduces the scale of the simulation pattern (including only the PI of specific nodes), enables us to swiftly obtain the simulation signatures of these nodes using exhaustive patterns.

Consider a scenario where a node possesses multiple fan-out connections, totaling $n$ in number. 
In the context of simulating all nodes, this specific node is accessed a total of $n+1$ times: once for computing its output and $n$ additional times for extracting its value.
By mapping these multiple fan-out nodes into a $k$-LUT, we effectively reduce unnecessary accesses. 
Consequently, we employ the STP to compute the TTs for each defined cut, employing nodes requiring simulation as the boundary. 
This strategic approach allows the simulator to efficiently acquire the desired node's value while incurring minimal computational overhead.

\subsection{Example}
\label{sec34}
\begin{figure}[t]
  \centering  
  \subfigbottomskip=-4pt 
  \subfigcapskip=-5pt 
  \subfigure[example logic circuit]{
      \label{fig:sim1}
      \includegraphics[width=0.29\textwidth]{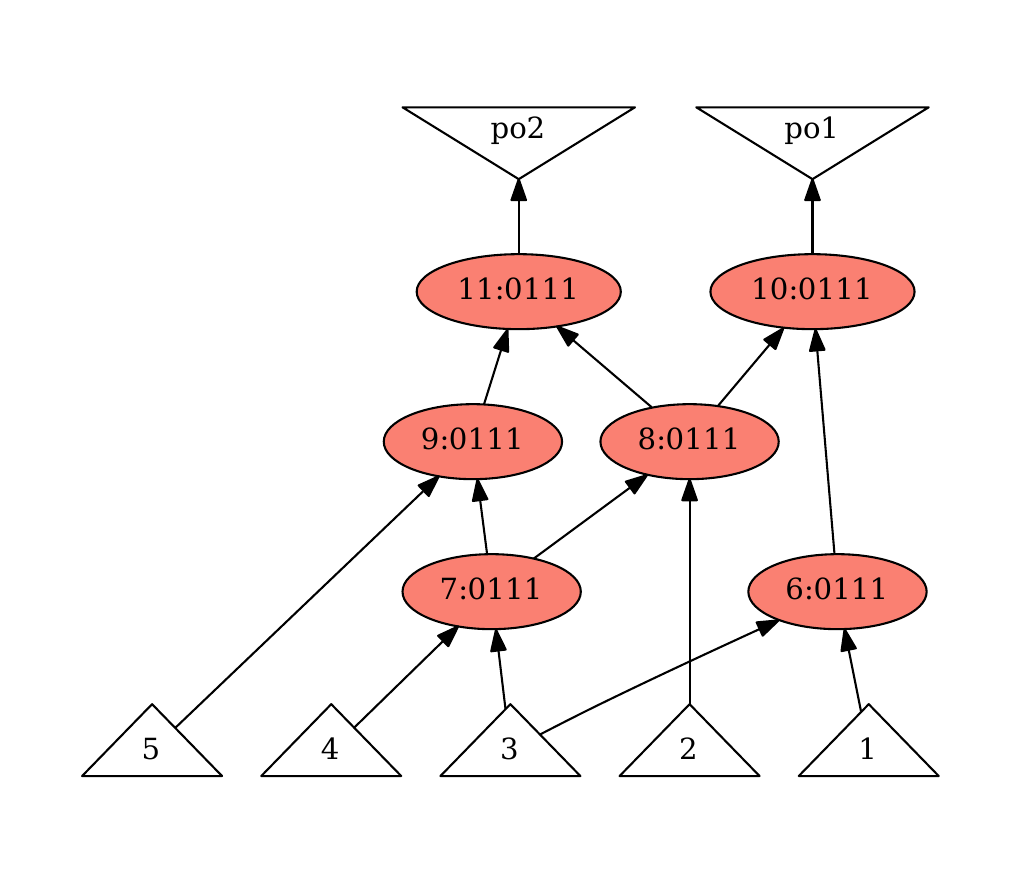}}
      \hspace{15mm}
  \subfigure[cutting algorithm]{
      \label{fig:sim2}
      \includegraphics[width=0.29\textwidth]{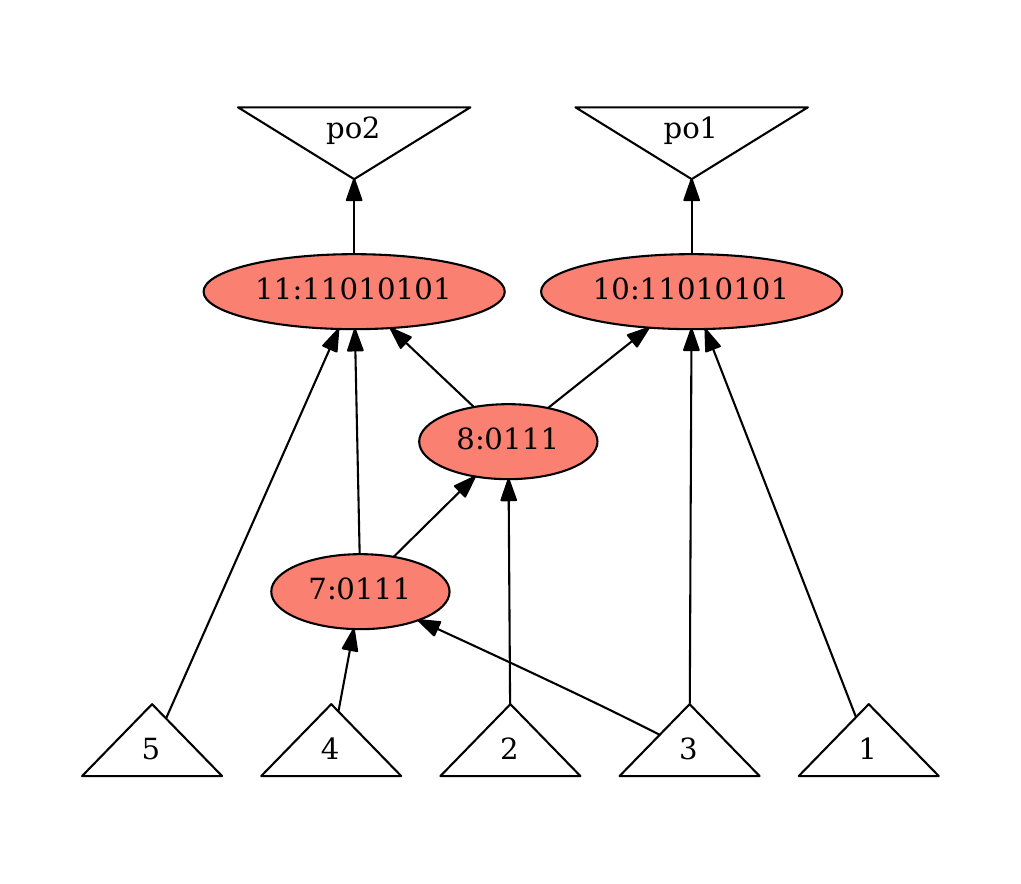}}
  \caption{Illustration of STP-based simulation}
  \label{fig:sim}
  \vspace{-2mm}
\end{figure}

Here we get a DAG which havs five PIs ($1, 2, 3, 4, 5$) and two POs ($po1, po2$), as shown in Fig.~\ref{fig:sim1}.
There are six intermediate nodes in the circuit, and each intermediate node records its TT. 
For example, the node ``6'' has two inputs ``1'' and ``3'' and one output ``10'', where the TT ``0111'' indicates that when the two inputs are assigned values according to the sequence, respectively ``11'', ``10'', ``01'', ``00'', the corresponding values of the output are ``0'', ``1'', ``1'', ``1'', that is, 2-input NAND.

Assume that there are 10 simulation patterns as
\begin{center}
$01110010111010011011111001100000000111111010000101,$
\end{center}
where the combination formed by taking out the $i$-th bit of each input represents the $i$-th simulation pattern.
For example, the first simulation pattern is ``01100''.

If there are two specific nodes, nodes ``7'' and ``8'' fall under this category.
We simulate and only get the simulation signature of these two nodes.
Firstly, we need to find the cut that satisfies the leaf node restriction. 
Since $3 < log_2^{10} < 4$, we set a limit of 3, meaning that each cut's input count should not exceed 3.
After cutting, we derive four distinct cuts: (6,10), (7), (8), and (9,11). 
For each cut, we proceed to compute its TT, as illustrated in Fig.~\ref{fig:sim2}.
Finally, all nodes of the root set are sorted according to the inverse topology order of the graph. 
We sequentially traverse and simulate each node within this sorted root set, ultimately leading to the acquisition of simulation signatures.
As an example, nodes ``7'' and ``8'' contain three PIs, namely ``2'', ``3'', and ``4''.
The scale of exhaustive patterns of node ``7'' and ``8'' is $2^2=4$, $2^3=8$, that is less than original 10 patterns.
By combining exhaustive patterns of PIs with the TT of node ``7'' (0111) and ``8'' (0111), the TT of node ``7'' and ``8'' can be simulated as follows: 7: 1110, 8: 11110001.

\vspace{-2mm}
\section{STP-based SAT-sweeping framework}
\label{sec4}
\begin{algorithm}[t]\footnotesize
  \SetAlgoLined
  \LinesNumbered 
  \caption{STP-based SAT-sweeping algorithm}
  \label{algo2}
  \KwIn{Network $N$, number $n$}
  \KwOut{Optimized network $N'$}
  $n \leftarrow 1000$ \;
  $S_e,S_c \leftarrow$ SAT\_guided\_simulation\_patterns($N$)\;
  \emph{class} $\leftarrow$ constant\_prop($N,S_c$), init\_equiv\_class($N,S_e$)\;
  \emph{list}  $\leftarrow$ inverse\_topo\_sort($N$)\;
  \ForEach{gate $G_i \in$ \emph{list} }{
    candidate = $G_i$\;
    \If{skip(candidate)}{
        \textbf{continue} \;
    }
    \emph{class\_new} = \emph{class}($G_i$) $\bigcup$ (INV + \emph{class}($\bar{G_i}$))\;
    sort\_topo\_order(\emph{class\_new}) \;
    \ForEach{gate $G_j \in$ \emph{class\_new}}{
        \ForEach{$G_k \in$ transitive\_fanin($G_j,n$)}{
            driver = $G_k$\;
            \If{skip(driver, candidate)}{
                \textbf{continue} \;
            }
            eq = SAT(candidate $\oplus$ driver)\;
            \If{eq == unDET}{
                mark\_dont\_touch(candidate)\;
                \textbf{break} \;
            }   
            \eIf{eq == unSAT}{
                $N' \leftarrow$ substitute\_node($G_j$, candidate)\;
            }{
                CE $\leftarrow$ SAT\;
                STP\_simulation($N$)\;
                refine\_equiv\_class(CE, class\_new, $N$)\;
            }
        }
    }
  }
  \textbf{return} $N'$\;
\end{algorithm}

\begin{figure}[h]
  \begin{center}
  \includegraphics[width=0.45\textwidth]{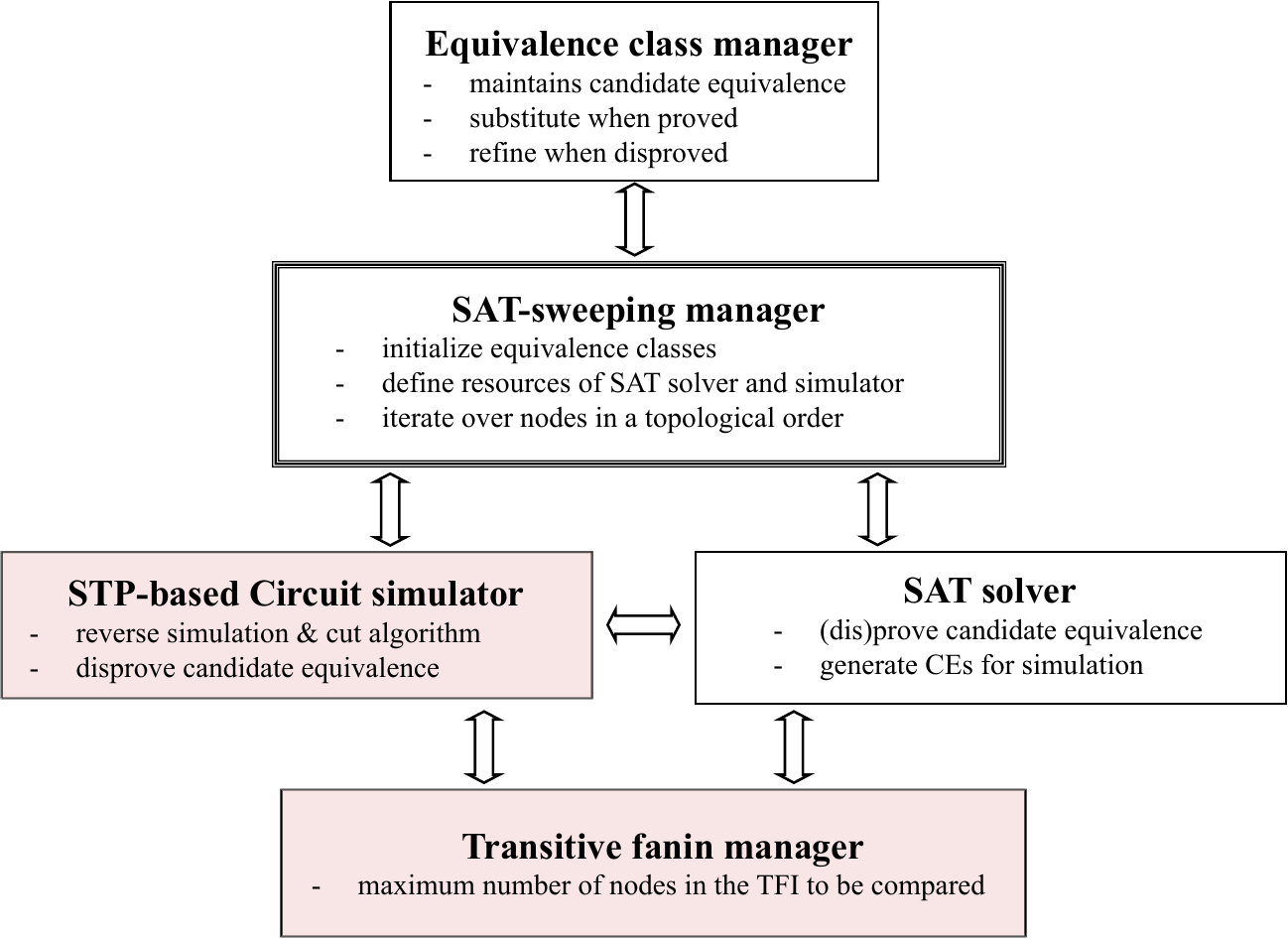}
  \caption{The proposed ecosystem.} 
  \label{fig:system}
  \end{center}
\end{figure}

In this section, we integrate our STP-based simulator into the SAT-sweeping framework, as depicted in Fig.~\ref{fig:system}, and the algorithm is delineated in Algorithm~\ref{algo2}.
In essence, our STP-based simulator accomplishes exhaustive simulation by employing circuit cut algorithm and focusing exclusively on simulating nodes within equivalence classes.
For the SAT solver, we utilize a circuit-based SAT solver to direct access to the network~\cite{sweep3}.

As for details, we set the upper limit for the number of nodes that can be compared within the TFI to 1000 (line 1). 
Simulation patterns are initially generated using SAT-guided initial pattern algorithms (line 2).
These high-quality simulation patterns serve as the foundation for computing equivalence classes, accounting for complementation, and executing the propagation of constant nodes for substitution (line 3).
Next, we arrange the list of gates that require processing in a reverse topological order, effectively traversing the circuit from primary outputs to primary inputs (line 4). 
Subsequently, we address the sorted gates in sequence. 
The gate presently under consideration is denoted as \emph{candidate}, as it is a potential candidate for removal and replacement with a preceding gate in the topological order (line 6).
To expedite the process, we promptly examine whether the current candidate should be skipped by checking for \emph{don't touch} conditions (lines 7-9). 
We treat equivalence classes for both positive and negative polarity as a single class, subsequently organizing this general class topologically (lines 10-11).
To maximize the QoR, we consider the TFI cones of each candidate to identify opportunities for valid merges (lines 12-13). 
As each member of the generalized equivalence class is attempted for a merge, it is called \emph{driver} (line 14).
Furthermore, it is essential to verify the conditions of \emph{drivers} (lines 15-17).
The equivalence problem is translated into \emph{Conjunctive Normal Form} (CNF) and submitted for resolution to the SAT solver (line 18). 
In the event that the solution is \emph{unDET}, signifying an undetermined outcome, the \emph{candidate} is marked as \emph{don't touch}.
This designation greatly enhances runtime speed and scalability. 
Subsequently, when we receive the \emph{unSAT}, we proceed with node substitution by connecting the fanins of the \emph{candidate} fanouts to the driver.
Upon receiving the \emph{SAT}, we retrieve the CE from the solver, employ the STP-based simulator to propagate the CE, and subsequently refine the equivalence classes based on this information. 
At the conclusion of this procedure, any dead nodes are eliminated from consideration.

\subsection{Refinement using STP-based Simulation}
\label{sec41}
A two-round SAT-guided simulation is first employed to perform tasks such as engine allocation, computation of candidate equivalence classes, and the substitution of constant nodes. 
The first round simulation is to ensure that gates exhibit simulation signatures characterized by either all zeros or all ones, which helps reduce network complexity through efficient constant propagation. 
Additionally, the second round simulation aims to avoid gates with only a few ones and the rest zeros, that is, simulation signatures with a high toggle rate\footnote{the toggle rate is the ratio of bit-toggles over the bit-string length.}.
The constraints on the characteristics of simulation patterns can be efficiently formulated as a SAT problem.
When the SAT problem is satisfied, new assignments will be generated at the PIs which satisfy the set of constraints~\cite{intro7}.

\vspace{-4mm}

CE simulation has been employed for the purpose of refining candidate equivalence classes. 
However, due to the impractical time demands of simulations, they often rely on partial simulation patterns.
To further enhance the refinement of candidate equivalence classes, STP-based simulation exclusively visits nodes belonging to the same equivalence class. 
For instance, if an equivalence class comprises only two nodes, ``A'' and ``B'', we map the remaining nodes into $k$-LUTs, and then simulate the values of ``A'' and ``B'' to disprove their equivalence.
Furthermore, during the simulation of each CE, we first convert nodes not within equivalence classes into $k$-LUTs, and then simulate candidate nodes to refine the equivalence classes. 
In contrast to traditional simulations, STP-based simulations adopt exhaustive simulation patterns, if the simulation is restricted to a window encompassing fewer than 16 leaf nodes.
In terms of practical implementation, the ID of the equivalence class is stored as an integer array within the \emph{equivalence class manager}.
This information is utilized by the STP-based simulator to exclusively simulate nodes belonging to the same equivalence class, while the remainder are mapped to $k$-LUTs.

\vspace{-2mm}
\section{Experimental results}
\label{sec6}
This section shows the effectiveness of the proposed circuit simulator and corresponding SAT-sweeping algorithm.
The proposed simulator is implemented in C++ on top of the logic synthesis framework ALSO\footnote{Chu Z. ALSO: Advanced logic synthesis and optimization tool. https://github.com/nbulsi/also, 2022.}, in which the source codes are publicly available. 
All experiments are performed on a 3.20 GHz Apple M1 CPU with 8GB of main memory.

\subsection{Simulation}
\label{sec61}
First, we evaluate the efficiency of our proposed STP-based simulator on the EPFL benchmarks suite\footnote{EPFL benchmark suite, https://github.com/lsils/benchmarks} by command `\texttt{simulator}' in ALSO.
We compare our proposed simulator (\emph{STP}) with the logic circuit simulation (\emph{Mockturtle}~\cite{exp0}).
Each benchmark is simulated with randomly generated $10^6$ simulation patterns.
\begin{table}[h]
	\centering
    \fontsize{8}{10}\selectfont
	\caption{Circuit simulation results for EPFL benchmarks.}
  \label{tab1}
	\begin{tabular*}{\hsize}{@{}@{\extracolsep{\fill}}lrrrrrr@{}}
    \toprule
		\multirow{2}{*}{Benchmark} & \multicolumn{2}{c}{\emph{Mockturtle}} & \multicolumn{4}{c}{\emph{STP}} \cr
    \cmidrule(l){2-3} \cmidrule(l){4-7}
    & $T_A$(s) & $T_L$(s) & $T_A$(s) & \texttt{x} & $T_L$(s) & \texttt{x} \cr
    \midrule
    adder & 1.33  & 12.56  & 1.36  & 0.98  & 0.57  & {\bf 22.04}   \\ 
    bar & 3.65  & 17.61  & 2.87  & 1.27  & 1.65  & 10.67   \\ 
    div & 65.12  & 414.92  & 69.04  & 0.94  & 56.30  & 7.37   \\ 
    hyp & 236.61  & 1,066.65  & 291.85  & 0.81  & 134.90  & 7.91   \\ 
    log2 & 35.30  & 181.67  & 36.01  & 0.98  & 45.17  & 4.02   \\ 
    max & 3.58  & 33.96  & 2.94  & 1.22  & 5.05  & 6.73   \\ 
    multiplier & 29.89  & 149.81  & 31.20  & 0.96  & 23.72  & 6.32   \\ 
    sin & 5.89  & 31.37  & 6.16  & 0.96  & 7.01  & 4.47   \\ 
    sqrt & 28.21  & 154.44  & 22.03  & 1.28  & 20.68  & 7.47   \\ 
    square & 20.42  & 88.91  & 24.26  & 0.84  & 11.34  & 7.84   \\ 
    artbiter & 12.91  & 67.43  & 14.99  & 0.86  & 21.02  & 3.21   \\ 
    cavlc & 0.77  & 3.09  & 0.81  & 0.95  & 0.53  & 5.83   \\ 
    ctrl & 0.21  & 0.77  & 0.17  & 1.24  & 0.09  & 8.58   \\ 
    dec & 0.28  & 4.98  & 0.62  & 0.45  & 0.90  & 5.54   \\ 
    i2c & 1.61  & 11.58  & 1.45  & 1.11  & 1.55  & 7.47   \\ 
    int2float & 0.28  & 1.47  & 0.28  & 1.00  & 0.18  & 8.19   \\ 
    mem\_ctrl & 52.21  & 312.26  & 49.74  & 1.05  & 54.02  & 5.78   \\ 
    priority & 1.19  & 10.03  & 0.95  & 1.26  & 1.17  & 8.58   \\ 
    router & 0.35  & 3.02  & 0.33  & 1.06  & 0.41  & 7.36   \\ 
    voter & 16.28  & 92.69  & 15.99  & 1.02  & 8.55  & 10.84  \\ 
  \midrule
  {\bf Geo.}& 4.99  & 30.44  & 5.04  & ~ & 4.24  &   \\ 
  {\bf Imp.}& 1.0   & 1.0    & {\bf 0.99}  & ~ & {\bf 7.18}  & ~ \\ 
  \bottomrule
	\end{tabular*}
  \begin{tablenotes}
    \footnotesize
    \item[1] Imp. : Average Geometric Mean Improvement (old/new).
  \end{tablenotes}
\end{table}
\begin{table*}[t]
  \centering
  \fontsize{8}{10}\selectfont
  \begin{threeparttable}
  \caption{Comparing the number of SAT calls and the runtime of the SAT sweepers (runtime in seconds)}
  \label{tab2}
  \begin{tabular*}{\hsize}{@{}@{\extracolsep{\fill}}lcrrrrrrrrrrrr@{}}
  \toprule
  \multirow{2}{*}{Benchmark} & \multicolumn{4}{c}{Statistics} & \multicolumn{2}{c}{SAT calls} & \multicolumn{2}{c}{Total SAT calls} & \multicolumn{2}{c}{Simulation} & \multicolumn{3}{c}{Total runtime} \cr
  \cmidrule(l){2-5} \cmidrule(l){6-7} \cmidrule(l){8-9} \cmidrule(l){10-11} \cmidrule(l){12-14} 
  & PI/PO &Lev & Gate  & Result & \&fraig & STP & \&fraig & STP& \&fraig & STP & \&fraig & STP & \texttt{x}  \\ 
  \midrule
  6s100        & 127138/97599  & 79      & 636,637    & 627,550   & 2,400   & 160    & 10,085  & 7,845  & 0.30   & 0.63  & 1.73    & 1.35   & 0.78   \\ 
  6s20         & 250/202       & 2,828   & 30,251     & 12,741    & 791     & 45     & 5,613   & 4,867  & 0.08   & 0.20  & 33.35   & 28.92  & 0.87   \\ 
  6s203b41     & 80192/68958   & 65      & 474,322    & 463,531   & 213     & 15     & 5,535   & 5,337  & 0.22   & 0.43  & 1.11    & 1.07   & 0.96   \\ 
  6s281b35     & 268334/177236 & 121     & 2,076,248  & 2,058,408 & 10,908  & 1,148  & 17,308  & 7,548  & 2.69   & 3.56  & 6.84    & 3.98   & 0.58   \\ 
  6s342rb122   & 59253/56839   & 52      & 330,130    & 319,365   & 194     & 9      & 3,224   & 3,039  & 0.06   & 0.19  & 0.24    & 0.23   & 0.94   \\ 
  6s350rb46    & 245680/243400 & 194     & 1,550,412  & 1,545,667 & 163     & 82     & 3,479   & 3,398  & 0.75   & 1.21  & 1.72    & 1.68   & 0.98   \\ 
  6s382r       & 106395/104831 & 2,752   & 1,756,654  & 1,704,409 & 1,158   & 74     & 5,911   & 4,827  & 1.87   & 4.08  & 86.36   & 70.52  & 0.82   \\ 
  6s392r       & 80920/80151   & 538     & 1,599,275  & 1,583,824 & 583     & 51     & 2,878   & 2,346  & 0.40   & 0.54  & 0.80    & 0.65   & 0.82   \\ 
  beemfwt4b1   & 3851/1595     & 1,212   & 47,368     & 41,580    & 1,009   & 222    & 1,412   & 625    & 0.28   & 0.68  & 16.28   & 7.21   & 0.44   \\ 
  beemfwt5b3   & 7370/3047     & 2,235   & 104,771    & 90,611    & 3,303   & 302    & 3,745   & 744    & 1.63   & 2.48  & 37.25   & 7.40   & {\bf 0.20}   \\ 
  oski15a07b0s & 8640/4454     & 3,822   & 120,268    & 118,728   & 3,812   & 953    & 4,073   & 1,214  & 2.47   & 3.38  & 17.06   & 5.08   & 0.30   \\ 
  oski2b1i     & 26489/13254   & 12,340  & 176,605    & 176,553   & 12,073  & 4,829  & 12,125  & 4,881  & 10.90  & 13.82  & 105.72  & 42.56  & 0.40   \\ 
  b18          & 6630/3322     & 64      & 131,100    & 124,530   & 606     & 17     & 3,926   & 3,337  & 0.05   & 0.25  & 0.40    & 0.34   & 0.85   \\ 
  b19          & 13235/6627    & 77      & 256,503    & 242,499   & 1,283   & 88     & 8,359   & 7,164  & 0.14   & 0.28  & 0.65    & 0.56   & 0.86   \\ 
  leon2        & 298888/291880 & 58      & 789,647    & 787,972   & 1,977   & 92     & 2,898   & 1,013  & 1.55   & 3.63  & 5.11    & 4.79   & 0.94   \\ 
  \midrule
  {\bf Geo.} & ~ & ~ & 335,363  & 305,780  & 1,230  & 119  & 4,943  & 2,965  & 0.53  & 1.04  & 4.83  & 3.15  &   \\ 
  {\bf Imp.} & ~ & ~ & 1.0  & {\bf 0.91}  & 1.0  & {\bf 0.09}  & 1.0  & {\bf 0.60}  & 1.00 & {\bf 1.99}  & 1.0  & {\bf 0.65}  & ~\\ 
  \bottomrule
  \end{tabular*}
  \begin{tablenotes}
    \footnotesize
    \item[1] Imp. : Average Geometric Mean Improvement (new/old).
    \end{tablenotes}
  \end{threeparttable}
  \end{table*}

The experimental results, as summarized in Table~\ref{tab1}, provide valuable insights. 
To facilitate comparison, we present the mean simulation time for both the \emph{And-Inverter Graph} (AIG), denoted as ``$T_A$'', and the $6$-LUT networks, denoted as ``$T_L$''. 
Additionally, we quantify the extent of CPU time acceleration as ``$\texttt{x}$''.
To gauge algorithm efficiency, we utilize the geometric mean, denoted as ``{\bf Geo.}'', and the average geometric mean improvement of optimized simulation over initialization, expressed as ``{\bf Imp.}''. 
It is evident that there is a preference for algorithms capable of achieving superior benchmark simulation times.
In terms of ``$T_A$'', our method demonstrates runtime performance comparable to that of \emph{Mockturtle}, which highlights the effectiveness of our method in simulating common data structures.
For ``$T_L$'', the \emph{STP} exhibits an average runtime improvement of 7.18$\times$ (22.04$\times$ maximum). 
These findings underscore the enhanced performance of our proposed simulator in the context of $k$-LUT simulation.
This advantage arises because \emph{Mockturtle} employs incremental simulation, which accelerates simulation by selectively re-simulating necessary nodes and re-computing only the last block of TT. 
However, it currently lacks support for $k$-LUT simulation. 
In $k$-LUT simulation, most simulators are limited to extracting individual bits of the LUT and simulating them separately.

\subsection{SAT-sweeping}
\label{sec62}
We implemented our method to demonstrate the effectiveness of the proposed SAT sweeper. 
Our experiments were conducted using a selected subset of benchmarks from the HWMCC'15~\cite{exp1} and IWLS'05\cite{exp2} benchmark suites.
The proposed SAT sweeper is based on existing engine, \emph{\&fraig} (command `\texttt{\&fraig -x}' in ABC\footnote{Mishchenko A. ABC: System for sequential logic synthesis and formal verification. https://github.com/berkeley-abc/abc, 2022}). 
As of our knowledge, \emph{\&fraig} represents the most efficient and scalable publicly available SAT sweeper.
All results are veriﬁed by `\texttt{\&cec}' command in ABC to ensure functional correctness.

As shown in Table~\ref{tab2}, we present the number of PIs and POs, internal AND-nodes in the original AIG (Gate), logic levels (Lev), and the internal AND-nodes after SAT-sweeping (Result) in the ``Statistics'' section.
Notably, the number of Result (9\% improvement of Gate) remains consistent across both engines since they start from the same AIG, with the SAT solver conflict limit disabled in all runs.
In ``SAT calls'' and ``Total SAT calls'' sections, we detail the number of satisfiable SAT calls and Total SAT calls, respectively.
We reduced the number of unsatisfiable SAT calls by using fast $k$-LUT based exhaustive simulations (restricted to a window encompassing fewer than 16 leaf nodes), thereby reducing the number of satisfiable solver calls by efficient equivalence class candidates.
Section ``Simulation'' shows the runtime for simulation, primarily encompasses initial simulation and CE simulation.
Lastly, the ``Total runtime'' section provides a direct comparison of the SAT sweeper's overall runtime, revealing a substantial 35\% reduction when compared to \emph{\&fraig}. 
While \emph{\&fraig} invests runtime resources in high-quality initial simulation, our method combines this with an exhaustive CE simulation. 
We exhibit almost 2$\times$ increase in simulation time, However, this is offset by a remarkable 91\% reduction in the number of SAT calls.
This hybrid approach not only improves the quality of simulation patterns but also dramatically reduces the number of SAT calls, ultimately resulting in highly efficient CE simulations.

\vspace{-2mm}
\section{Conclusion}
\label{sec7}
In this paper, we introduce a novel STP-based circuit simulation tailored for SAT-sweeping. 
Simulating $k$-LUT networks can be particularly challenging for conventional fast bitwise simulators available off-the-shelf. 
STP, however, provides advantages in the computation of logic matrices and circuit connectivity.
The primary contribution of this paper lies in the introduction of cut algorithm that substantially enhances simulation speed and promotes efficient information reuse. 
When compared to the SOTA simulator, our STP-based simulation demonstrates an average reduction in CPU time by an average of 7.18$\times$.
Moreover, our simulation approach is deeply integrated into the SAT sweeper, leading to substantial reductions in runtime for refining equivalence classes. 
Consequently, the implemented SAT sweeping approach, on average, performs 35\% reduction in runtimes, without compromising quality.


\begin{thebibliography}{}
\providecommand{\url}[1]{#1}
\csname url@samestyle\endcsname
\providecommand{\newblock}{\relax}
\providecommand{\bibinfo}[2]{#2}
\providecommand{\BIBentrySTDinterwordspacing}{\spaceskip=0pt\relax}
\providecommand{\BIBentryALTinterwordstretchfactor}{4}
\providecommand{\BIBentryALTinterwordspacing}{\spaceskip=\fontdimen2\font plus
\BIBentryALTinterwordstretchfactor\fontdimen3\font minus
  \fontdimen4\font\relax}
\providecommand{\BIBforeignlanguage}[2]{{%
\expandafter\ifx\csname l@#1\endcsname\relax
\typeout{** WARNING: IEEEtran.bst: No hyphenation pattern has been}%
\typeout{** loaded for the language `#1'. Using the pattern for}%
\typeout{** the default language instead.}%
\else
\language=\csname l@#1\endcsname
\fi
#2}}
\providecommand{\BIBdecl}{\relax}
\BIBdecl

\end{thebibliography}


\footnotesize
\begin{thebibliography}{00}
    \bibitem{intro1} Mishchenko A, Brayton R. ``Integrating an AIG package, simulator, and SAT solver,'' in \emph{Proc.IWLS}, 2018: 11-16.
    \bibitem{intro3} Mishchenko A, Chatterjee S, Brayton R, et al. ``Improvements to combinational equivalence checking,'' in \emph{Proc.ICCAD}, 2006: 836-843.
    \bibitem{intro4} Zhu Q, Kitchen N, Kuehlmann A, et al. ``SAT sweeping with local observability don't-cares,'' in \emph{Proc.DAC}, 2006: 229-234.
    \bibitem{intro2} Zhang H T, Jiang J H R, Amarú L, et al. ``Deep integration of circuit simulator and SAT solver,'' in \emph{Proc.DAC}, 2021: 877-882.
    \bibitem{intro6} Cheng D, Qi H, Xue A. ``A Survey on Semi-Tensor Product of Matrices,'' in \emph{Journal of Systems Science Complexity}, 2007, 20: 304-322.
    \bibitem{intro7} Amarú L, Marranghello F, Testa E, et al. ``SAT-sweeping enhanced for logic synthesis,'' in \emph{Proc.DAC}, 2020: 1-6.
    \bibitem{pri1} Lee SY, Riener H, Mishchenko A, et al. ``A simulation-guided paradigm for logic synthesis and verification,'' \emph{IEEE Transactions on Computer-Aided Design of Integrated Circuits and Systems}, 2021, 41(8): 2573-2586.
    \bibitem{pri2} Cheng D, Qi H, Zhao Y. ``An introduction to semi-tensor product of matrices and its applications,'' \emph{World Scientific}, 2012.
    \bibitem{pri21} Pan HY and Chu ZF, ``Exact Synthesis Based on Semi-Tensor Product Circuit Solver,'' in \emph{Proc.DATE}, 2023, pp. 1-6, doi: 10.23919/DATE56975.2023.10137287.
    \bibitem{pri3} Van Loan C F. ``The ubiquitous Kronecker product,'' \emph{Journal of computational and applied mathematics}, 2000, 123(1-2): 85-100.
    \bibitem{pri4} Mishchenko A, Chatterjee S, Jiang R, et al. ``FRAIGs: A unifying representation for logic synthesis and verification,'' \emph{ERL Technical Report}, 2005.
    \bibitem{sweep2} Mishchenko A, Zhang J S, Sinha S, et al. ``Using simulation and satisfiability to compute flexibilities in Boolean networks,'' \emph{IEEE Transactions on Computer-Aided Design of Integrated Circuits and Systems}, 2006, 25(5): 743-755.
    \bibitem{sweep1} Lu F, Wang L C, Cheng K T, et al. ``A circuit SAT solver with signal correlation guided learning,'' in \emph{Proc.DATE}, 2003: 892-897.
    \bibitem{sweep3} Zhang H T, Jiang J H R, Mishchenko A. ``A circuit-based SAT solver for logic synthesis,'' in \emph{Proc.ICCAD}, 2021: 1-6.
    \bibitem{exp0} Siang-Yun  L. mockturtle: a head-only library for logic synthesis and logic optimization, https://github.com/lsils/mockturtle.
    \bibitem{exp1} HWMCC'15 benchmarks https://fmv.jku.at/hwmcc15/.
    \bibitem{exp2} IWLS 2005 benchmarks https://iwls.org/iwls2005/benchmarks.html.
\end{thebibliography}
\end{document}